# Disclosing the network structure of private companies on the web: the case of Spanish IBEX 35 share index

Enrique Orduña-Malea*
*EC3 Research Group. Institute of Design and Manufacturing (IDF). Polytechnic University of Valencia.*

Emilio Delgado López-Cózar
*EC3 Research Group. Facultad de Comunicación y Documentación. Universidad de Granada.*

Jorge Serrano-Cobos
Nuria Lloret-Romero
*Institute of Design and Manufacturing (IDF). Polytechnic University of Valencia.*

* Corresponding author: Camino de Vera s/n, Valencia 46022, Spain. E-mail: enorma@upv.es

## Abstract
**Purpose**
It is common for an international company to have different brands, products or services, information for investors, a corporate blog, affiliates, branches in different countries, etc. If all these contents appear as independent additional web domains (AWD), the company should be represented on the web by all these web domains, since many of these AWDs may acquire remarkable performance that could mask or distort the real web performance of the company, affecting therefore on the understanding of web metrics. The main objective of this study is to determine the amount, type, web impact and topology of the additional web domains in commercial companies in order to get a better understanding on their complete web impact and structure.

**Design/methodology/approach**
The set of companies belonging to the Spanish IBEX-35 stock index has been analyzed as testing bench. We proceeded to identify and categorize all AWDs belonging to these companies, and to apply both web impact (web presence and visibility) and network metrics.

**Findings**
The results show that AWDs get a high web presence but relatively low web visibility, due to certain opacity or less dissemination of some AWDs, favoring its isolation. This is verified by the low network density values obtained, that occur because AWDs are strongly connected with the corporate domain (although asymmetrically), but very weakly linked each other.

**Research limitations/implications**
The categories used to classify the various AWD, although they are clearly distinguishable conceptually, have certain limitations in practice, since they depend on the form adopted by companies to publish certain content or to provide certain services or products. Otherwise, the use of web indicators presents certain problems of accuracy that could be softened if applied with caution and in a relational basis.

**Originality/value**
Although the processes of AWDs creation and categorization are complex (web policy seems not to be driven by a defined or conscious plan), their influence on the web performance of IBEX 35companies is meaningful. This research measures the AWDs influence on companies under webometric terms for the first time.

## 1. Introduction

The application of web indicators on the performance of companies through cybermetric techniques is relatively low when compared with other applied areas of this discipline, especially those studies focused on universities (Aguillo, Granadino, Ortega, & Prieto,

2006), academic journals (Vaughan & Thelwall, 2003) and, more recently, alternative metrics (Thelwall, 2014). However, during the last decade a limited but growing research line whose object of study are the commercial companies has been developing (Romero-Frias, 2011).

We can mainly distinguish a first area focuses on discovering the possible existence of significant correlations between web indicators, especially web visibility (number of links that target to a particular web domain) of companies' corporate websites, with financial variables (Vaughan, 2004), and second area based on the study of the benefits of certain web indicators, especially those that can be used to understand the relationships between companies in homogeneous and heterogeneous sectors (Vaughan & Wu, 2004). Particularly noteworthy are the studies performed from co-links (websites A and B are co-linked "x" times when we found "x" websites that provide a link both to A and B) (Vaughan & You, 2006) and co-words (words A and B are co-worded "x" times when we found "x" websites that provide a mention both to word A and B) (Vaughan & You, 2010).

In parallel there have been numerous studies based on the web platforms likely to provide the necessary data to construct indicators. For example, general search engines, such as Alltheweb, Altavista, MSN and Google (Vaughan & Wu, 2004) and Yahoo! (Vaughan & You, 2006); specific search engines, like Google News and Google Blogs (Vaughan & Romero-Frias, 2012); and web traffic sources, such as Alexa, Compete and Google Trends (Vaughan & Yang, 2013).

These works expanded the application of cybermetrics outside academic environments (where the objects of study are both academic contents or academic-related agents, such as universities, journals, authors, etc.), exploiting web data for competitive intelligent purposes (Reid, 2003), under the assumption that link and web content analysis can be useful both as a technique to gathering soft information (non-statistical intelligence gathered from non-traditional sources) in general, as to identify competitor's relationships (official or unofficial) in particular.

Nonetheless, the application of web metrics to commercial companies has a number of important shortcomings. For example, the analysis of direct links between firms is not recommended since companies rarely tend to generate links to their competitors (Shaw, 2001). Moreover they do not publish certain information because of trade secrets, and their websites' impact (web visibility, web traffic, etc.) may depend on many external factors (stock sale and purchase of businesses movements, event sponsorship, etc.), difficult to identify and correct, and making their web impact widely distributed and complex to measure accurately (Thelwall, 2000; 2001).

One of the biggest limitations is determined by the complexity of company websites, due not only to their content structure (determined by parameters of information architecture and interaction design), but on the existence of multiple web domains in which contents are driven, that reflect indirectly the real structure of these firms.

This complexity is determined not only by the existence of sub-domains or sub-folds (which obviously exist), but by the existence of multiple web domains, technically independent of the official corporate website, intended to provide targeted information directly or indirectly related to the company.

In this sense, it is common for an international company to have different brands, specific products or services, information websites for investors, a corporate blog, affiliates or branches in different countries (with its own products, services, etc., acting as a fractal), etc. If all these websites appear as separate web domains or additional web domains (AWD hereinafter), the company should be represented on the web by every AWD, and not just by the corporate web domain.

While many of these AWDs surely get a minimum web impact, some others may acquire remarkable performance (perhaps even more than the corporate web domain) that could mask or distort the real web performance of the company, affecting therefore on the understanding of the web indicators used.

To date, research on the application of web indicators to companies have been based solely on the usage of corporate web domains, avoiding the various existing AWDs. Therefore, the main objective of this study is to determine the amount, type, web impact and topology of the additional web domains in commercial companies, in order to get a better understanding on the complete web impact and structure of these companies. The specific objectives are the following:

(1) Identify, quantify and categorize additional web domains in a sample of international trading companies.
(2) Measure the contribution of the additional web domains on the web impact (measured by page count and web visibility) of the companies.
(3) Shape the complete company network by analyzing the degree of interconnectivity between all additional web domains.

## 2. Related research

Although some pioneering works (Wormell, 2001; Tan, Foo, & Hui, 2002) had already used link analysis for competitive intelligence, it is not until the publication of the classic works of Vaughan and Wu (2003) and Vaughan (2004) when cybermetric indicators started systematically to be applied in business environments. These early works were promising as they found certain web indicators, especially external inlinks (links to a website that come from pages not on the same root domain) correlating significantly with some financial variables (revenue, profit, and research and development expenses), even if external variables such as firm size or website age were uncontrolled, proving thus that inlinks contain useful business information.

It was subsequently discovered that these correlations were higher when the group of companies is homogeneous (Vaughan & Wu, 2004) and local (Vaughan, Gao, & Kipp, 2006), depending also on the industrial sector (some business activities attract more links than others, especially those industries whose business model is more information-centered and consumer-oriented). Thereby telecommunications or financing sectors perform better than chemical (Vaughan, Yang, & Tang, 2012) or heavy construction (Vaughan & Romero-Frías, 2012).

Additionally, some financial variables (total assets, revenue and net income) shown to be especially correlated with external links since all of them are of cumulative nature (Romero-Frías and Vaughan, 2010). Recently, web traffic data have demonstrated its usefulness to estimate online business performance as well (Vaughan and Yang, 2013).

Complex indicators such as co-links (Vaughan, Tang, & Du, 2009) and co-words (Vaughan, Yang, & Tang, 2012) were developed in parallel, providing useful tools to analyze online relationships. Advanced methods (such as the combination of co-links and keywords) were also introduced in order to examine products or services (Vaughan, & Yu, 2008; 2009). Dependence on geographic and linguistic factors (Romero-Frías, & Vaughan, 2010) and web data source (Vaughan & Romero-Frias, 2012) were also investigated.

However, under an informetric perspective, it is necessary to comprehend the different motivations that led to the creation of these links for a better understanding of the nature of these findings. Vaughan, Gao and Kipp (2006) confirm that the most common type of website linking to North American IT companies was commercial sites (71.3%). In the same way, Vaughan, Kipp and Gao (2007) found that 61.4% of co-links to IT companies were created to connected pairs of highly related businesses (related companies, related products, and related services).

The fact that these studies are applied only to company official websites, may favor the attraction of this business-related links, so that web hyperlinks built a network that connects businesses, customers, suppliers etc. (Vaughan, Gao, & Kipp, 2006). Nonetheless, additional web domains, focused on certain services or products and targeted to final users, may attract mentions to the company of different nature, which could affect the previous findings. Therefore, the study of the possible influence of AWDs justifies the development of this research.

## 3. Method

### 3.1. Collection and categorization of the sample

The sample corresponds to the set of companies belonging to the IBEX-35 stock index (Spanish Exchange Index), a market capitalization weighted index which comprises the 35 most liquid Spanish stocks traded in the Madrid Stock Exchange General Index.

The choice of this sample is motivated by the convenience to manage international companies with large budgets (therefore likely to have high web impact and large variety of additional web domains), belonging to different industrial sectors, and operating on a national index (being all governed by similar web legal requirements).

IBEX-35 share index has served indeed as a testing bench to diverse cybermetric studies, especially those performed by Guillamón-Saorín and Martínez-López (2013), focused on the reliability of corporate financial information on the web, and by Romero-Frias and Vaughan (2010), who analyze five stock Exchange: Dow Jones Industrial-30 (US), EuroStoxx-50 (Euro zone), CAC-40 (Paris), FTSE-100 (London), and IBEX-35 (Madrid), using the official corporate websites to carry out all web measures. This research can thus complement the former.

From the official website of the IBEX-35 at ibex35.com, the 35 companies listed in the index at the time of data collection (February 2014) were identified (see Appendix A), the main industrial sector category to which they belong were extracted as well, at http://www.bolsamadrid.es/esp/aspx/Empresas/Empresas.aspx.

Next, we manually proceeded to identify the URL of each company. A manual process of identifying AWDs was performed by analyzing all menus, sitemaps and outlinks. Only independent web domains were considered, avoiding subdomains and sub-folds. Afterwards, we proceeded to categorize all AWDs. The classification scheme was developed through an induction process obtaining 8 general categories (see Table 1). This process was conducted from February to April 2014.

**Table 1. Additional web domain categories**

*3.2. Web measures*

In Table 2 we show all indicators used as well as their scope, source, and the corresponding query performed to gather web data.

**Table 2. Web impact and network indicators**

The page count measure is obtained through Google by using the "site" command; Google's choice is motivated by being the commercial search engine with the most comprehensive index of the web today. Although the "site" command is not exhaustive, the possible error rate introduced is equally distributed to all web domains, so its effect is minimized if data are taken in a comparative manner. The use of an ad hoc web crawler may be a possible alternative, but this procedure is more time consuming, being only recommended when analyzing few web domains (Thelwall & Sud, 2011). Furthermore, the use of robot exclusion protocols may limit data gathering.

Web visibility indicators are obtained from Alexa, at alexa.com, and Open Site Explorer (OSE), at opensiteexplorer.com. In both cases we have opted for the use of site-level metrics (i.e., they quantify the amount of websites that link to one URL instead of quantifying the amount of external links). On the one hand we select the "sites linking in" indicator in Alexa, recommended by Vaughan and Yang (2012). On the other hand, we select the "Total Root Domains linking" from OSE, previously utilized by Orduna-Malea, Torres-Salinas & Delgado López-Cózar (in press). Finally, the "Domain Authority" is collected from OSE; this indicator reflects the reputation of each web domain in a similar fashion as PageRank does. Its use is recommended as its range (0 to 100) is more discriminating than the PageRank (0 to 10). All measures were conducted during June 2014.

Since there are no longer tools providing selective inlink data today (the number of links from a website A to website B), the inter-relationship between AWDs has been calculated with URL mentions, an indicator that has already been successfully tested as a substitute (Thelwall, Sud, & Wilkinson, 2012).

A query for all possible combinations between all AWDs of each company was automatically constructed. For example, the query <"acciona.com" site:acciona-engineering.com> returns the number of times the web domain <acciona.com> is mentioned in the web domain <acciona-engineering.com>, being therefore an indicator about the degree of connectivity between two web domains from the same company.

Only companies with at least 10 AWDs (26) were considered for network analysis (less nodes makes the network unusable by introducing mathematical artifacts to all network indicators, thus providing some unrealistic values). The queries for all combinations (31,250) were conducted via Google directly, through the Hit Count Estimates (HCE). The procedure was performed manually because Google does not provide API (Application Programming Interface). Google was selected due to its higher coverage and the inability of Bing API to obtain HCE for web domains over 1,000 hits (http://lexiurl.wlv.ac.uk).

The data were exported into a spreadsheet from which NET files were manually built for each company, and then exported to Gephi v. 0.8.2, from which we obtained directly both node-level indicators (Degree, Closeness, Betweenness, Eigenvector, Clustering coefficient) for each AWD, and network-level indicators (Average degree, Diameter, Density, Average clustering coefficient, Average path length), for each company (Table 2).

Finally, XLstat suite was used to perform the following statistical analyzes:

- Correlation analysis: web impact indicators and node-level indicators were correlated each other in order to find any possible relationship among them. Since web data presents a skewed distribution, Spearman ($\alpha=0.01$) was applied in all calculations.
- Principal Component Analysis (PCA): it was performed in order to complement correlation by finding causes that explain the variability of the indicators applied to the sample of AWDs. The Pearson (n) PCA with varimax rotation was applied.

## 4. Results

### 4.1. Additional web domains

Considering the 35 companies, a total of 818 AWDs have been identified, although their distribution per company is uneven ($\bar{x}= 23$; $\sigma= 19.7$). On the one hand, The Santander group (with 98 AWDs) and Telefónica (70) are the two companies with the most localized web domains, followed at some distance by OHL (41) and Ebro Foods (40). By contrast, only 1 AWD has been identified for REE, constituting thus a high statistical range (97).

All additional web domains identified in each of the 35 companies according to category are shown in Table 3.

**Table 3. Number of additional web domains according to category**

"Related" is the most numerous category (294), followed by "brands & products" (192), while "division" is scarce (36), and are mainly due to Acciona.

No specific pattern is observed in the distribution of AWDs according to category. It is noteworthy on the one hand a cluster formed by Santander and Telefónica (characterized by a large number of AWDs associated to "delegations", "brands & products" and "other services"), and on the other hand some special cases of companies that concentrate most of their AWDs in a particular category, such as "division" in Acciona, "service" in Mapfre, "delegation" in Gas Natural, and finally a larger group of companies with a concentration of "related company" (Sabadell, ACS, Sacyr, FCC, or BME).

### 4.2. Web impact

The page count results roughly show a high performance on "brands & products". In fact, of the 50 AWDs better ranked according to page count, up to 37 are product-oriented, especially noteworthy are those related with Telefónica (Movistar and Terra), Santander (Universia) and Inditex (Zara) companies.

Some "related company" AWDs can also be highlighted, such as Tele 5 ("telecinco.es") and Cuatro TV channels ("cuatro.com"), owned by Mediaset, or Iberia ("iberia.com") and British Airways ("britishairways.com"), which were merged into the IAG group. Service AWDs achieve significant results as well, particularly financial information services ("bsmarkets", "santandertrade.com") and real estates ("servihabitat.com" or "bbvavivienda.com"), offered by all banks.

By contrast, "corporate" AWDs show lower performances, among which LaCaixa ("lacaixa.es")*, Mapfre ("mapfre.com") and Bankia ("bankia.es"), all of which belong to the financial service and real estate sector, may be noted. These results, coupled with the large number of existing AWDs on some companies, explain the low share of corporate web domains on the total amount of pages that the company owns (Table 4).

**Table 4. Contribution of Corporative web domains on the complete count page of companies**

The "corporate" web domain page count represents less than 50% of the total page count on 20 of the 35 companies analyzed, being especially critical on Inditex, Mediaset, IAG, Santander, Banc Sabadell and Telefónica, in which the percentage is less than 1%. Otherwise, the Spearman correlation coefficient between "corporate" page count and total page count, although significant, does not achieve good results ($r = .452$; $\alpha=0.01$).

Web visibility metrics show meanwhile an improvement in the performance on "corporate" web domains ("arcelormittal.com", "repsol.com") and "delegation", for example Santander-Brazil ("Santander.com.br") and Telefónica-Spain ("telefonica.es"). Despite this, the AWD category with largest number of domains linking them are "brands and products" ("terra.com.br", "zara.com") and "related companies" ("britishairways.com", "iberia.com").

These results are nonetheless determined largely by the lack of data on web visibility of the sources used. In Alexa, the data obtained at 39 web domains is null, while this figure rises to 265 (32.4%) in OSE.

At this point, beyond the results at the individual level (the existence of companies with many powerful AWDs, such as Telefonica and Santander, can skew the results), is of great interest to know, for each company, which AWD category is generating the greater impact. To this end, Table 5 shows, for each company, the number of AWDs by category that achieve the best result in each of the four web impact indicators applied.

**Table 5. Number of additional web domains with the higher performance according to category for each web impact indicator**

The "corporate" web domain becomes the largest according to page count indicator only on 18 companies (51.4%), while "related company" achieves this on 8 companies, and "brands & products" on 4. Finally, also stands out the case of "guiarepsol.com", better-ranked web domain according to page count in Repsol, reflecting thus the importance of some complementary services on the business web impact.

In the case of web visibility metrics (according to Alexa) the "corporate" AWD constitutes the web domain type with the greater value in the 65.7% of companies, and a "related company" web domain is the better-ranked for 6 companies, confirming the importance of this AWD category.

*4.3. Additional web domains inter-relation*

All network-level indicators of each company are shown in Table 6.

**Table 6. Network level indicators for IBEX 35 companies (n=26)**

Companies with the highest average Degree are Telefónica (16.74) and Santander (13.33), which correspond to companies with more additional web domains (see Table 1), followed at some distance by Banc Sabadell (9.28). Broadly speaking, high diameters (average for the 26 companies considered is 3.5) and very low densities (average density is 0.2) are obtained. On the other hand, high average clustering coefficients (especially for SACYR, LaCaixa, Mapfre, Ferrovial, ACS and Jazztel, all of them exceed the value of "0.7"), are obtained. This, coupled with the low Average page length values (only 5 companies exceed the value of "2"), indicate a "small-world" effect in the companies.

If we shift the analysis from network indicators to node indicators, we observe an increase in the importance of corporate web domains. These are the most influential node in their respective network (measured by Eigenvector centrality) in 77% of the companies.

The importance of this AWD category is reflected equally both in Degree and Betweenness metrics. In Table 7 we display illustratively the 10 web domains with higher performance on these indicators. These data also reveal a high web impact of various delegations (Spain and Germany) and trademarks (Movistar) in Telefonica network. Likewise, Closeness metric complements this showing the absence of corporate web domains with higher closeness values.

**Table 7. Best performers on InDegree, OutDegree, Betweenness, and Closeness for IBEX 35 companies (n=26)**

However, disaggregating Degree indicator into inDegree and outDegree, we can observe a high asymmetry between these two metrics for some nodes. Both positive (authority) and negative (hub) nodes are identified (the asymmetry is greater than "10" in 7.3% of 772 nodes).

While nodes with more authority seem to correspond with delegations, products and services, hub nodes contain a higher number of corporate web domains. In fact, only two corporate AWDs have a positive asymmetry above of "10" ("gasnatural.com", 14; "bbva.es", 12), while up to 10 have a negative asymmetry under "-10". The nodes with higher asymmetries are presented in Table 8, where a hub behavior for some corporate web domains, such as Ferrovial ("ferrovial.com", "ferrovial.es") or ACS ("grupoacs.com") is appreciated.

**Table 8. Network indicators. Asymmetry between InDegree and OurDegree for IBEX 35 companies (n=26)**

In any case, it should be noted that Degree metric quantifies the number of nodes to which a connection is established, but does not quantify the intensity of this connection, i.e., the number of mentions established between two nodes. Thus we have calculated the total number of URL mentions from a corporate node to all non-corporate nodes (acting as source, thereby becoming hubs) and vice versa (acting as target, becoming authorities).

The results do not show clear patterns of behavior (Table 9). Some corporate web domains are mainly targets of mentions ("bbva.com", "santander.com"), while others are sources (especially "telefonica.com").

**Table 9. URL mention among corporative web mention for IBEX 35 companies and combination better connected**

Additionally, in Table 9 we show for each corporate web domain the combination of nodes with a greater intensity among them. Noteworthy are the relationships obtained between "santander.com.mx" and "casacompara.com.mx" (a Mexican delegation and related company of Santander group), "pullandbear.net" and "pullandbear.com" (two existing URLs of the same brand of Inditex), and finally "terra.com" and "terra.com.br" (a Telefonica's brand and a delegation of such brand), although in this case, the accuracy of the indicator is jeopardized due to the URL string syntax (See Discussion section).

Finally, and as special case studies, the complete maps for the two companies with the greatest number of additional web domains (Telefonica and Santander) are presented (Figures 1 and 2).

**Figure 1. Telefonica network (n= 70; Force-directed graph drawing algorithm: Fruthterman Reingold)**
**Figure 2. Santander network (n= 98; Force-directed graph drawing algorithm: Fruthterman Reingold)**

In the case of Telefónica, despite the great centrality of its corporate web domain ("telefonica.com"), a concentration of nodes is observed for two brands (Movistar and Terra), which have a major set of additional web domains (mainly delegations), producing a fractal effect within the network with its own corporate and delegation web domains. As regards Santander, this fractal effect is also observed for Universia brand, formed by a large set of nodes with high Degree and clustering coefficient.

*4.4. Web impact and Network indicators correlation*

The correlation data between all indicators used are available in Table 10, where you can see how the web impact indicators correlated significantly ($\alpha = 0.01$), except for a few special exceptions, with network node-level indicators.

**Table 10. Correlation matrix among web impact and network indicators**

Regarding web impact indicators, a high correlation between web visibility indicators is obtained (Alexa shows a correlation of .94 with OSE, and .90 with Authority metrics). On the other hand, the page count also correlates with web visibility indicators (r=.67 with Alexa, and r=.66 with OSE), albeit at a slightly lower level.

As regards to the relationship between web impact and network metrics, the page count achieves moderated correlation with InDegree (r=.58) and OutDegree (r=.65). The web visibility indicators obtain similar correlations, though weaker, with OutDegree (r=.4).

The Principal Component Analysis (Figure 3) graphically shows the relationships between the various metrics. Thus, a gap between the web impact and network indicators is observed. Moreover, within these two dimensions, a slight separation between the page count and web visibility (in the case of web impact), and between the Eigenvector and InDegree, Outdegree and Betweenness (for network indicators) is observed as well. Finally, the clustering coefficient and closeness represent independent dimensions of the rest.

**Figure 3. Principal Component Analysis (PCA) of web impact and network indicators**

**5. Discussion**

The categories used to classify the various additional web domains, although they are clearly distinguishable conceptually, have certain limitations in practice, since they depend on the form adopted by companies to publish certain content or to provide certain services or products.

Therefore, unwanted circumstances occur. For example, "Servihabitat" and "BankiaHabitat" are both real estate services offered by two banks. However, in the former case is considered "service", and the second a "related" company. The reason depends on whether the website clearly informed of the existence of a distinct legal entity associated with this web domain or not. This does not happen in the first case (although the legal entity may exist, but not reported in this website). This problem specifically affects the distinction among "related", "Brand & Product" and "Service" categories.

This creates an interpretation problem of the data provided in Tables 3 and 5. Therefore, it is recommended to follow these three categories as one more general category, since other categories are clearly distinguishable themselves. In any case, the corporate web domain isolation from the rest of the AWDs is the distinction that must be taken into account.

Similarly, subjectivity with which these additional web domains are created must be considered, as many companies offer the same content, but structured under sub-folds. For example, FCC has both services ("fcc.es/servicios") and divisions ("fcc.es/infraestructuras") in the form of sub-folds, which do not represent independent web domains.

The clearest example are "Delegations", presenting a broad and diverse casuistic in the URL syntaxes. For example, some companies use first level subdirectories as Repsol ("repsol.com/pe_es/") or multilevel, as Grifols ("grifols.com/es/web/Portugal"), not constituting thus additional web domains.

However, other companies use randomly different approaches. The most complex case is Amadeus, which are detected up to 4 different techniques, where only the first one constitutes an additional web domain:
   (1) Using a ccTLD ("amadeus.cl");
   (2) Use of specific sub-domain ("br.amadeus.com");
   (3) Use of 2-level sub-folds ("amadeus.com/ro/x41234.xml");
   (4) Use of sub-folds of more than 2 levels ("amadeus.com/web/amadeus/es_EC-EC/Amadeus-Home/1259102412927-Page-AMAD_HomePpal")

Apart from the structure that online information adopts and its categorization, the use of web indicators presents certain problems that should be discussed.

All measures based on web indicators should be taken with some caution, in particular those obtained through the search commands of Google, due to the high variability of the data and a number of inconsistencies, recently summarized by Willinson and Thelwall (2013). All measures based on Hit Count Estimates (especially Page count and URL mentions) should be treated cautiously since Google provides only rounded values. These limitations especially affect the use of these indicators to evaluate web performance (e.g. the exact number of mentions to a URL). However, if used for relational purposes (e.g. to determine whether the relationship between "website 1" and "website 2" is greater than that between "website 1" and "website 3", which is how it should be interpreted in this study), these limitations are minimised since all URLs are subject to the same error, and thus the error is statistically dispersed.

In the case of web visibility (Alexa and OSE), these results must be taken with caution too, as they are unable to discriminate the source of the links. Therefore, in each AWD, links that may be coming from other AWDs are included. In any case, the fact that these indicators work at site-level minimizes this effect.

Furthermore, it should be noted the absence of data for a large number of AWDs. In Alexa, 32 AWDs have a visibility equal to "0", and no information is available for other 39. This figure is higher in OSE, where no data are available for 265 AWDs (32.4% of all the analyzed AWDs), although their effect in the correlations obtained in Table 10 is estimated to be reduced, since they constitute web domains with scarce web impact.

For net indicators, a lack of precision in the use of URL mentions in those web domains where the AWD syntax contains the entire corporate web domain, for example <"**acciona.com**" site:**acciona.com**.br>. This means that if the string "acciona.com.au" is detected inside "acciona.com.br" website, Google will account this result as a mention of "acciona.com", when it is not the case.

These data directly affect the intensity of the relationships obtained in section 4.3, and are especially significant for Terra, because of the nomenclature of their delegations. The following apparently inflated results are observed:

<"terra.com" site:**terra.com**.br>: 11,800,000 mentions; (query 1)
<"terra.com" site:**terra.com**.pe>: 2,320,000 mentions; (query 2)
<"terra.com" site:**terra.com**.mx>: 2,220,000 mentions; (query 3)
<"terra.com" site:**terra.com**.co>: 1,240,000 mentions; (query 4)
<"terra.com" site:**terra.com**.ar>: 1,220,000 mentions. (query 5)

One possible solution would be to add a blank space after the mention: "terra.com [space]", but this option is not properly interpreted by Google, so it should not be used.

Another option is to manually delete the results from other queries that can potentially affect. For example, in the case of "**terra.com**.br", would be subtracted from the total figure (11,800,000), the results of the following queries:

<"terra.com.ar" site:**terra.com**.br>: 19 results;
<"terra.com.co" site:**terra.com**.br>: 57 results;
<"terra.com.ec" site:**terra.com**.br>: 6 results;
<"terra.com.sv" site:**terra.com**.br>: 1 results;
<"terra.com.mx" site:**terra.com**.br>: 36 results;
<"terra.com.pe" site:**terra.com**.br>: 29 results;
<"terra.com.ve" site:**terra.com**.br>: 12 results;

This involves removing only 160 results. However, in the remaining domains, results are larger. In the corresponding case of "terra.com.pe" this figure rises to 123,875 results, and keeps high on "terra.com.mx" (89,974), "terra.com.co" (110,323) and "terra.com.ar" (57,773). Nonetheless, despite these raw results are high, the percentages related to the total mentions gathered from the general queries 1-5 are reduced, so that the influence of other delegations in the results is limited.

In fact, it is the influence of the web domain that acts as "source" which distorts the data. That is, the query <"terra.com" site:**terra.com**.br> retrieves all results in which is mentioned "terra.com.br" thereby distorting the results. Moreover, if the query <"terra.com.br" site:**terra.com**.br> is applied, virtually the same results are obtained. It is therefore impossible to isolate the entries corresponding to "terra.com" with those corresponding to "terra.com.br". Moreover, as Google only returns the first 1,000 results per query, a manually filtering process is not even possible.

This fact must be checked in Figures 1 and 2, in relation to the Terra (Telefónica) and Universia (Santander) brands. So these graphs show, for this type of relationship between delegations, what Google search commands are able to discriminate and not the actual mentions. Fortunately, this problem affects only to measures related to the intensity of relations measured by URL mentions, and not to other network-level measures.

In any case, URL mentions present problems not only on a technical level, but also on the company web culture. The report "Presence of Ibex 35 companies in the Web 2.0", produced by *Estudio de comunicación* (2013), indicated that most of IBEX 35 companies had trouble finding a style of communication and patterns of behavior that matched in its corporate culture and they were effective in the social web as well.

This different communication style involves online content generation patterns different from other non-financial environments. In most cases, for example, instead of incorporating a hyperlink to a company or product (from an AWD or from another company), a textual mention without hypertext associated is used. However, both synonymy and polysemy, and the existence of different acronyms for each company make impossible to quantify precisely these mentions, although some studies have already applied them (Vaughan and Romero-Frias, 2012), by using the full name of companies (called Title mentions or web keywords) in specific cases.

Finally, it is necessary to discuss about the greater or lesser need to quantify these different AWDs. Strictly speaking, a link targeted to an additional web domain belongs to the company, though the corporate website has not received this link.

However, there are some products whose impact is rarely associated with the corresponding company. For example, Heathrow Airport is a product of Ferrovial, but most links to "heathrow.com" or "heathrowairport.com" are rarely made thinking about Ferrovial. And this happens frequently in construction companies (whose products are buildings, airports, highways, etc.).

Another important situation is given by the purchase and sale of companies or mergers. For example, links to British Airways and Iberia, should we really associate them with IAG? And the links to the social network "tuenti.com", should it be assigned to Telefónica? In these cases, a large number of links were generated probably without having in mind the parent company, or probably having in mind the previous owner company.

## 6. Conclusions

Given the potential limitations discussed in the previous section, the results show a large complexity of additional web domains associated with the IBEX-35 companies. While some companies such as Santander and Telefónica have a large number and variety of AWDs, other companies only a few AWDs are detected, showing cultures and approaches to the Web completely different, and where the business sector does not seem to be influencing.

The most used AWD category corresponds to "related" company. Up to 74.3% of firms have at least one AWD belonging to this category. By contrast, the category "Division" is the least used, with only 17.1% of companies using it. In any case, a great randomness was observed in the presence of certain categories. In some companies a high use of some AWD categories (e.g. delegations in Natural Gas, Divisions in Acciona or Services in Mapfre) is performed while AWDs for other activities are not used. Therefore, the choice of using either subdomains or sub-folds seems not to be driven by a defined or conscious policy.

In any case, the page count corresponding to the corporate web domain accounts for only a small percentage of the total pages that depend on the company (if we also consider the "site" command is not exhaustive, this percentage is likely to be greater). Moreover, the correlation between corporate page count and the total page count is low (r= .45). Thus, it is found that the application of page count indicators only to corporate web domains is not representative of the actual size of these companies.

By contrast, web visibility metrics show a higher impact of corporate web domains. This may be due to certain opacity or less dissemination of some AWDs. Moreover, the fact of constituting independent web domains involves loss of heritage reputation (in terms of PageRank) of the corporate web domain, which could be minimized with high connectivity in terms of links between the corporate domain and the rest of AWDs.

Notwithstanding, network analysis has shown, with some exceptions, the existence of high diameters and low densities, which reflect a certain isolation of many AWDs, thus losing web visibility. Likewise, the asymmetries between corporate web domains and other AWDs show a lack of reciprocity in the network of web links within the company, favoring the isolation of the AWDs.

By contrast, small groups of AWDs highly connected among themselves and with high clustering coefficients (generally formed by corporate web domain and its Delegations or Services) are identified, in which a small-world effect is detected as well.

Regarding the intensity of relationships, the empirical analysis has shown the inefficiency of using "URL mentions" in Google to study the relationships between web domains that share the URL syntax. Saving this circumstance, the data support that the most influential nodes on all networks are the corporate web domains.

The different types of indicators used, both web impact (page count and web visibility) and network (Degree, Betweenness), correlate significantly with each other, though measuring complementary dimensions, as the PCA shows, reflecting the need to use all them to get a more complete picture of companies.

But the motivations for linking each AWD are obviously different. The corporate web domains are those that offer company stock information as well as information of interest to investors so that they attract some specific business-related links. So maybe only these web domains should be used to search for correlations with financial indicators.

This work therefore represents just an exploratory study of the influence of the AWDs on companies' websites, when they are analyzed through cybermetric indicators. Although the processes of AWDs creation and categorization are complex, their influence on the web performance of companies has been established (especially in page count).

Finally, aspects such as the motivations behind hyperlinks to each type of AWD (which hypothetically may vary depending on their category), the evolution of indicators over time (more or less fluctuation, dependence on market movements, news releases, etc.), or the possible influence of the age of the AWDs (how long they need to achieve their current web impact) are parameters to be analyzed in future works to extend our understanding of companies' web structures measured by webometric indicators.

Likewise, the incorporation of new web indicators, especially usage indicators (e.g., visits), is estimated necessary. This work is focused intensively in the links and mentions between URLs; however it is possible that very popular (visited) AWDs are not linked with the parent company website (and vice versa). Therefore, the future incorporation of these indicators would help further understanding of the influence of the AWDs in web companies.

**Table 1.**
Additional web domain categories

| CATEGORY | SCOPE |
| --- | --- |
| **CORPORATIVE** | Official website of the company domain, which contains the corporate and financial information |
| **DELEGATION** | Web domain corresponding to a territorial delegation |
| **RELATED** | Web domain corresponding to an affiliate or associate to main company |
| **BRAND AND PRODUCT** | Web domain corresponding to a particular product or brand developed and / or offered by the company |
| **DIVISION** | Web domain corresponding to a specific section, department or line of work of the company |
| **SERVICE** | Web domain corresponding to a service offered by the company, both B2C and B2B |
| **FOUNDATION** | Web domain corresponding to a foundation associated with or linked directly to web domain studied. |
| **OTHERS** | Web domains not included in the above categories, especially services not directly related to the activities of the company: blogs, information services, event sponsorship, selling tickets to cultural events, museums, training facilities, etc. |

**Table 2**
Web impact and network indicators

| WEB IMPACT | SCOPE | SOURCE | QUERY |
|---|---|---|---|
| **Page count** | Number of files indexed within a web domain | Google | site:abc.com |
| **Sites linking in** | Number of unique root domains containing | Alexa | Direct |
| **Total root domains linking** | at least one link to an specific URL | OSE | Direct |
| **Authority** | Web domain popularity score (0 to 100) | OSE | Direct |
| **URL mention** | Number of times a one URL is mentioned from another specific URL | Google | "abc.com" site:xyz.com |

| NODE LEVEL | SCOPE |
|---|---|
| **Degree** | Number of edges that are adjacent to one node |
| **Closeness** | How often a node appears on shortest paths between nodes in the network |
| **Betweenness** | The average distance from a given node to all other nodes in the network |
| **Clustering coefficient** | The degree to which nodes of the neighborhood of a node "a" are connected each other |

| NETWORK LEVEL | SCOPE |
|---|---|
| **Average degree** | The average degree over all of the nodes in the network |
| **Average clustering coefficient** | The average clustering coefficient over all of the nodes in the network |
| **Average path length** | The average graph-distance between all pairs of nodes |
| **Diameter** | The maximal distance between all pairs of nodes |
| **Density** | How close the network is to complete (density equal to 1) |

**Table 3.**
Number of additional web domains according to category

| COMPANY | COR | DEL | REL | BRA | DIV | SER | FOU | OTH | TOTAL | SECTOR |
|---|---|---|---|---|---|---|---|---|---|---|
| **Santander** | 1 | 13 | 34 | 28 | 0 | 7 | 3 | 12 | **98** | Financial and real estate services |
| **Telefónica** | 1 | 14 | 3 | 38 | 0 | 0 | 1 | 13 | **70** | Technology and Communications |
| **OHL** | 1 | 2 | 17 | 18 | 3 | 0 | 0 | 0 | **41** | Basic materials, industry and construction |
| **Ebro Foods** | 1 | 0 | 15 | 23 | 0 | 0 | 1 | 0 | **40** | Consumer services |
| **Acciona** | 1 | 6 | 6 | 0 | 25 | 0 | 0 | 1 | **39** | Basic materials, industry and construction |
| **Ferrovial** | 2 | 0 | 16 | 19 | 0 | 0 | 0 | 0 | **37** | Basic materials, industry and construction |
| **Mapfre** | 1 | 3 | 10 | 0 | 0 | 21 | 1 | 1 | **37** | Financial and real estate services |
| **Banc Sabadell** | 4 | 3 | 20 | 1 | 0 | 5 | 0 | 3 | **36** | Financial and real estate services |
| **Abertis** | 1 | 0 | 7 | 25 | 0 | 0 | 1 | 1 | **35** | Consumer services |
| **Inditex** | 3 | 0 | 0 | 25 | 0 | 0 | 0 | 2 | **30** | Consumer goods |
| **ACS** | 1 | 0 | 27 | 0 | 0 | 0 | 0 | 0 | **28** | Basic materials, industry and construction |
| **Iberdrola** | 3 | 3 | 16 | 2 | 0 | 0 | 1 | 2 | **27** | Oil & Energy |
| **Sacyr** | 3 | 0 | 17 | 7 | 0 | 0 | 0 | 0 | **27** | Basic materials, industry and construction |
| **Gas Natural** | 2 | 18 | 3 | 0 | 0 | 0 | 1 | 1 | **25** | Oil & Energy |
| **Caixabank** | 3 | 0 | 13 | 0 | 2 | 6 | 0 | 0 | **24** | Financial and real estate services |
| **FCC** | 1 | 0 | 23 | 0 | 0 | 0 | 0 | 0 | **24** | Basic materials, industry and construction |
| **BBVA** | 2 | 0 | 0 | 0 | 3 | 5 | 4 | 9 | **23** | Financial and real estate services |
| **BME** | 1 | 0 | 17 | 0 | 0 | 1 | 0 | 0 | **19** | Financial and real estate services |
| **IAG** | 1 | 0 | 8 | 0 | 0 | 2 | 0 | 8 | **19** | Consumer services |
| **Arcelor Mittal** | 1 | 4 | 10 | 0 | 2 | 0 | 0 | 0 | **17** | Basic materials, industry and construction |
| **BPE** | 1 | 2 | 4 | 1 | 0 | 6 | 0 | 2 | **16** | Financial and real estate services |
| **Mediaset** | 1 | 0 | 11 | 2 | 0 | 0 | 0 | 2 | **16** | Consumer services |
| **Bankia** | 2 | 0 | 4 | 0 | 0 | 5 | 1 | 0 | **12** | Financial and real estate services |
| **DIA** | 6 | 5 | 0 | 0 | 0 | 0 | 0 | 0 | **11** | Consumer services |
| **Grifols** | 2 | 1 | 3 | 0 | 0 | 0 | 3 | 2 | **11** | Consumer goods |
| **Jazztel** | 3 | 0 | 0 | 0 | 0 | 6 | 0 | 1 | **10** | Technology and Communications |
| **Enagás** | 5 | 0 | 0 | 3 | 0 | 0 | 0 | 0 | **8** | Oil & Energy |
| **Repsol** | 4 | 0 | 1 | 0 | 0 | 0 | 1 | 2 | **8** | Oil & Energy |
| **Bankinter** | 2 | 0 | 1 | 0 | 0 | 2 | 1 | 0 | **6** | Financial and real estate services |
| **Gamesa** | 1 | 0 | 5 | 0 | 0 | 0 | 0 | 0 | **6** | Basic materials, industry and construction |

| | COR | DEL | REL | BRA | DIV | FOU | OTH | | Total | Sector |
|---|---|---|---|---|---|---|---|---|---|---|
| **Viscofan** | 1 | 1 | 3 | 0 | 1 | 0 | 0 | 0 | **6** | Consumer goods |
| **Amadeus** | 2 | 3 | 0 | 0 | 0 | 0 | 0 | 0 | **5** | Technology and Communications |
| **Indra** | 1 | 3 | 0 | 0 | 0 | 0 | 0 | 0 | **4** | Technology and Communications |
| **Técnicas Reunidas** | 2 | 0 | 0 | 0 | 0 | 0 | 0 | 0 | **2** | Basic materials, industry and construction |
| **REE** | 1 | 0 | 0 | 0 | 0 | 0 | 0 | 0 | **1** | Oil & Energy |
| **TOTAL** | **68** | **81** | **294** | **192** | **36** | **66** | **19** | **62** | **818** | |

COR: Corporative; DEL: Delegation; REL: Related companies; BRA: Brands and products; DIV: Division; FOU: Foundation; OTH: Others.

**Table 4.**
Contribution of Corporative web domains on the complete count page of companies.

| COMPANY | COUNT PAGE | | |
|---|---|---|---|
| | CORPORATIVE | TOTAL | % |
| **Inditex** | 3,118 | 5,907,672 | **0.053** |
| **Mediaset** | 9,930 | 3,365,739 | **0.295** |
| **IAG** | 3,710 | 1,138,645 | **0.326** |
| **Santander** | 60,000 | 17,225,739 | **0.348** |
| **Banco Sabadell** | 4,337 | 958,701 | **0.452** |
| **Telefónica** | 243,000 | 46,292,475 | **0.525** |
| **Ebro Foods** | 2,800 | 76,757 | **3.648** |
| **DIA** | 2,002 | 35,682 | **5.611** |
| **BBVA** | 42,810 | 596,144 | **7.181** |
| **BPE** | 3,620 | 47,621 | **7.602** |
| **Ferrovial** | 25,660 | 204,404 | **12.554** |
| **Acciona** | 4,740 | 34,642 | **13.683** |
| **BME** | 66,600 | 442,142 | **15.063** |
| **FCC** | 24,600 | 121,667 | **20.219** |
| **SACYR** | 3,453 | 11,859 | **29.117** |
| **OHL** | 5,450 | 16,329 | **33.376** |
| **ACS** | 11,600 | 34,603 | **33.523** |
| **Repsol** | 168,250 | 423,734 | **39.707** |
| **Abertis** | 13,700 | 29,893 | **45.830** |
| **Bankia** | 317,300 | 674,285 | **47.057** |
| **Viscofan** | 2,710 | 4,420 | **61.312** |
| **Iberdrola** | 45,060 | 59,919 | **75.202** |
| **Gas Natural** | 88,962 | 114,270 | **77.852** |
| **Caixabank** | 1494,759 | 1,889,036 | **79.128** |
| **Grifols** | 5,600 | 6,797 | **82.389** |
| **Mapfre** | 559,000 | 654,638 | **85.391** |
| **Gamesa** | 5,670 | 6,382 | **88.844** |
| **Arcelor Mittal** | 44,200 | 49,530 | **89.239** |
| **Bankinter** | 68,299 | 74,801 | **91.308** |
| **Indra** | 37,600 | 38,927 | **96.591** |
| **Enagás** | 33,617 | 34,735 | **96.781** |
| **Jazztel** | 31,309 | 31,581 | **99.139** |
| **Amadeus** | 285,000 | 2,850,48 | **99.983** |
| **REE** | 6,790 | 67,90 | **100.000** |
| **Técnicas Reunidas** | 4,420 | 44,20 | **100.000** |

**Table 5.**
Number of additional web domains with the higher performance according to category for each web impact indicator

| CATEGORY | Page Count | Alexa | OSE | Authority | TOTAL |
|---|---|---|---|---|---|
| **CORPORATIVE** | 18 | 23 | 21 | 22 | **84** |
| **DELEGATION** | 2 | 2 | 3 | 2 | **9** |
| **VINCULATED** | 8 | 6 | 7 | 7 | **28** |
| **BRAND & PRODUCT** | 4 | 3 | 4 | 3 | **14** |
| **DIVISION** | 0 | 0 | 0 | 0 | **0** |
| **SERVICE** | 2 | 0 | 0 | 0 | **2** |
| **FOUNDATION** | 0 | 0 | 0 | 0 | **0** |
| **OTHERS** | 1 | 1 | 0 | 1 | **3** |

**Table 6.**
Network level indicators for IBEX 35 companies (n=26)

| COMPANY | AVERAGE DEGREE | DIAMETER | DENSITY | AVERAGE CLUSTERING COEFF. | AVERAGE PATH LENGH |
|---|---|---|---|---|---|
| **Telefónica** | **16.743** | 4.000 | 0.243 | 0.593 | 1.849 |
| **Santander** | **13.337** | 5.000 | 0.537 | 0.560 | 2.750 |
| **Banco Sabadell** | **9.278** | 4.000 | 0.265 | 0.667 | 1.837 |
| **Gas Natural** | **7.640** | 3.000 | 0.318 | 0.696 | 1.722 |
| **Iberdrola** | **7.111** | 4.000 | 0.274 | 0.645 | 1.750 |
| **BME** | **6.947** | 2.000 | 0.386 | 0.633 | 1.596 |
| **IAG** | **5.895** | 5.000 | 0.327 | 0.524 | 1.952 |
| **SACYR** | **5.667** | 3.000 | 0.218 | 0.768 | 1.699 |
| **Acciona** | **5.231** | 4.000 | 0.138 | 0.576 | 1.958 |
| **Mediaset** | **5.062** | 3.000 | 0.338 | 0.645 | 1.822 |
| **Caixabank** | **4.833** | 4.000 | 0.210 | 0.705 | 1.901 |
| **BBVA** | **4.826** | 3.000 | 0.219 | 0.607 | 1.769 |
| **Inditex** | **4.667** | 4.000 | 0.161 | 0.564 | 2.263 |
| **Mapfre** | **4.216** | 3.000 | 0.117 | 0.714 | 1.892 |
| **BPE** | **4.000** | 4.000 | 0.267 | 0.594 | 1.800 |
| **Bankia** | **3.917** | 3.000 | 0.356 | 0.675 | 1.788 |
| **Ferrovial** | **3.838** | 3.000 | 0.107 | 0.775 | 1.979 |
| **Abertis** | **3.800** | 4.000 | 0.112 | 0.390 | 2.137 |
| **ACS** | **2.929** | 3.000 | 0.108 | 0.712 | 1.738 |
| **OHL** | **2.683** | 4.000 | 0.067 | 0.542 | 2.110 |
| **FCC** | **2.625** | 2.000 | 0.114 | 0.468 | 1.849 |
| **Jazztel** | **2.600** | 3.000 | 0.289 | 0.757 | 1.765 |
| **Grifols** | **2.000** | 3.000 | 0.200 | 0.583 | 1.738 |
| **Ebro Foods** | **1.950** | 5.000 | 0.050 | 0.294 | 2.755 |
| **Arcelor Mittal** | **1.706** | 2.000 | 0.107 | 0.246 | 1.839 |
| **DIA** | **1.636** | 3.000 | 0.164 | 0.324 | 1.714 |

**Table 7.**
Best performers on InDegree, OutDegree, Betweenness, and Closeness for IBEX 35 companies (n=26)

| URL | INDEGREE | URL | BETWEENNESS |
|---|---|---|---|
| santander.com | 68 | santander.com | 3561.029 |
| santander.de | 67 | universia.net | 1870.198 |
| telefonica.com | 57 | telefonica.com | 1014.346 |
| telefonica.es | 55 | santanderconsumer.com | 870.586 |
| santander.no | 55 | santander.de | 618.846 |
| telefonica.de | 54 | mapfre.com | 615.667 |
| movistar.es | 49 | ohl.es | 541.633 |
| movistar.com | 45 | telefonica.es | 512.738 |
| universia.net | 42 | ferrovial.com | 449.500 |
| universia.es | 42 | acciona.es | 443.450 |
| **URL** | **OUTDEGREE** | **URL** | **CLOSENESS*** |
| universia.net | 77 | santanderconsumerusa.com | 3.191 |
| santander.com | 63 | roadloans.com | 3.213 |
| telefonica.com | 60 | lacigala.es | 3.294 |
| telefonica.com.pe | 46 | carolinarice.com | 3.394 |
| telefonica.com.ve | 46 | gourmethouserice.com | 3.455 |
| telefonica.es | 45 | watermaidrice.com | 3.455 |
| universia.es | 43 | minuterice.com | 3.485 |
| movistar.es | 42 | americanbeauty.com | 3.485 |
| universia.edu.uy | 39 | lightnfluffy.com | 3.485 |
| universia.net.co | 38 | chryslercapital.com | 4.149 |

* Closeness indicator is better as it is lower whereas Degree and Betweenness are better as they are higher.

**Table 8.**
Network indicators. Asymmetry between InDegree and OurDegree for IBEX 35 companies (n=26)

| URL | Positive asymmetry | URL | Negative asymmetry |
|---|---|---|---|
| **santander.de** | 57 | **telefonicachile.cl** | -20 |
| **santander.no** | 53 | **grupoacs.com** | -21 |
| **telefonica.de** | 41 | **dragados-usa.com** | -21 |
| **movistar.com** | 26 | **cesvimap.com** | -21 |
| **cervantesvirtual.com** | 22 | **ferrovial.com** | -22 |
| **becas-santander.com** | 21 | **tecsa-constructora.com** | -23 |
| **redemprendia.org** | 20 | **telefonica.com.pe** | -26 |
| **fundacionbancosantander.com** | 19 | **ferrovial.es** | -29 |
| **o2.com** | 19 | **universia.net** | -35 |
| **innoversia.net** | 17 | **telefonica.com.ve** | -42 |

**Table 9.**
URL mention among corporative web mention for IBEX 35 companies and combination better connected

| COMPANY | URL | TARGET | SOURCE | TOP MENTION COMBINATION | MENTION |
|---|---|---|---|---|---|
| **Abertis** | abertis.com | 1,044 | 34 | "autopistas.com" site:abertis.com | 655 |
| **Acciona** | acciona.com | 1,388 | 1,488 | "acciona.es" site:acciona.com | 1,060 |
| **ACS** | grupoacs.com | 304 | 2,232 | "grupocobra.com" site:grupoacs.com | 1,090 |
| **Arcelor Mittal** | arcelormittal.com | 1,184 | 187 | "arcelormittal.com" site:arcelormittal.kz | 483 |
| **BBVA** | bbva.com | 15,538 | 1,086 | "bbva.com" site:ligabbva.com | 21,600 |
| **BPE** | grupobancopopular.com | 34 | 891 | "bancopopular.es" site:grupobancopopular.com | 767 |
| **B Sabadell** | grupbancsabadell.com | 121 | 1,375 | "sabadellcam.com" site:bsandorra.com | 3,270 |
| **Santander** | santander.com | 70,351 | 41,581 | "santander.com.mx" site:casacompara.com.mx | 237,000 |
| **Bankia** | bankia.es | 1,362 | 2,147 | "bankia.es" site:bankia.com | 947 |
| **BME** | bolsasymercados.es | 268 | 3,937 | "bmemarketdata.es" site:borsabcn.es | 21,800 |
| **Caixabank (Lacaixa)** | lacaixa.es* | 9,845 | 5,951 | "lacaixa.es" site:caixabank.com | 6,870 |
| **DIA** | diacorporate.com | 21 | 144 | "diacorporate.es" site:dia.es | 8,940 |
| **Ebro Foods** | ebrofoods.es | 13 | 103 | "oryza.de" site:euryza.de | 57 |
| **Ferrovial** | ferrovial.com | 1,055 | 2,825 | "heathrow.com" site:heathrowairport.com | 46,500 |
| **FCC** | fcc.es | 333 | 196 | "fcc.es" site:fccenvironment.co.uk | 105 |
| **Gas Natural** | gasnaturalfenosa.com | 9,052 | 1,866 | "gasnaturalfenosa.com" site:gasnaturalfenosa.es | 2,080 |
| **Grifols** | grifols.com | 544 | 613 | "grifols.es" site:grifols.com | 541 |
| **IAG** | iairgroup.com | 646 | 1,121 | "iberia.es" site:iberia.com | 24,800 |
| **Iberdrola** | iberdrola.es | 5,883 | 4,279 | "iberdrola.com" site:iberdrola.es | 3,280 |
| **INDITEX** | inditex.com | 16,569 | 514 | "pullandbear.net" site:pullandbear.com | 676,003 |
| **Jazztel** | jazztel.com | 190 | 148 | "jazztel.es" site:jazztel.com | 131 |
| **Mapfre** | mapfre.com | 24,149 | 22,913 | "mapfre.es" site:mapfre.com | 16,500 |
| **Mediaset** | mediaset.es | 2,299 | 3,818 | "mitele.es" site:cuatro.com | 113,000 |
| **OHL** | ohl.es | 456 | 121 | "ohlconcesiones.com" site:ohlconcesiones.com.pe | 83 |
| **SACYR** | sacyr.com | 908 | 883 | "sacyr.com" site:gruposyv.com | 476 |
| **Telefónica** | telefonica.com | 41,040 | 430,559 | "terra.com" site:terra.com.br | 118,00,000 |

*Although Caixabank is the Company included in IBEX 35, we have considered La Caixa (mother entity) as reference.

**Table 10.**
Correlation matrix among web impact and network indicators

|       | Pco       | Alexa     | OSE       | Aut       | InD       | OutD      | Clo       | Bet       | Cco   | Eve  |
|-------|-----------|-----------|-----------|-----------|-----------|-----------|-----------|-----------|-------|------|
| **Pco**   | 1.00      |           |           |           |           |           |           |           |       |      |
| **Alexa** | **0.67**  | 1.00      |           |           |           |           |           |           |       |      |
| **OSE**   | **0.66**  | **0.94**  | 1.00      |           |           |           |           |           |       |      |
| **Aut**   | **0.68**  | **0.91**  | **0.94**  | 1.00      |           |           |           |           |       |      |
| **InD**   | **0.58**  | **0.53**  | **0.53**  | **0.58**  | 1.00      |           |           |           |       |      |
| **OutD**  | **0.65**  | **0.47**  | **0.46**  | **0.50**  | **0.77**  | 1.00      |           |           |       |      |
| **Clo**   | -0.08     | -0.05     | -0.05     | -0.07     | -0.10     | -0.00     | 1.00      |           |       |      |
| **Bet**   | **0.55**  | **0.54**  | **0.53**  | **0.54**  | **0.75**  | **0.80**  | -0.07     | 1.00      |       |      |
| **Cco**   | **-0.12** | **-0.22** | **-0.21** | **-0.20** | -0.07     | **-0.16** | **0.22**  | **-0.40** | 1.00  |      |
| **Eve**   | **0.41**  | **0.38**  | **0.39**  | **0.39**  | **0.73**  | **0.50**  | **-0.28** | **0.55**  | -0.03 | 1.00 |

\*\* significant values (except diagonal) at the level of significance alpha=0,010 (two-tailed test)
Pco: Page count; Aut: Authority; InD: InDegree; OutD: OutDegree; Bet: Betweenness; Cco: Clustering coefficient; Eve: Eigenvector

**Figure 1. Telefonica network (n= 70; Force-directed graph drawing algorithm: Fruthterman Reingold)**

**Figure 2. Santander network (n= 98; Force-directed graph drawing algorithm: Fruthterman Reingold)**

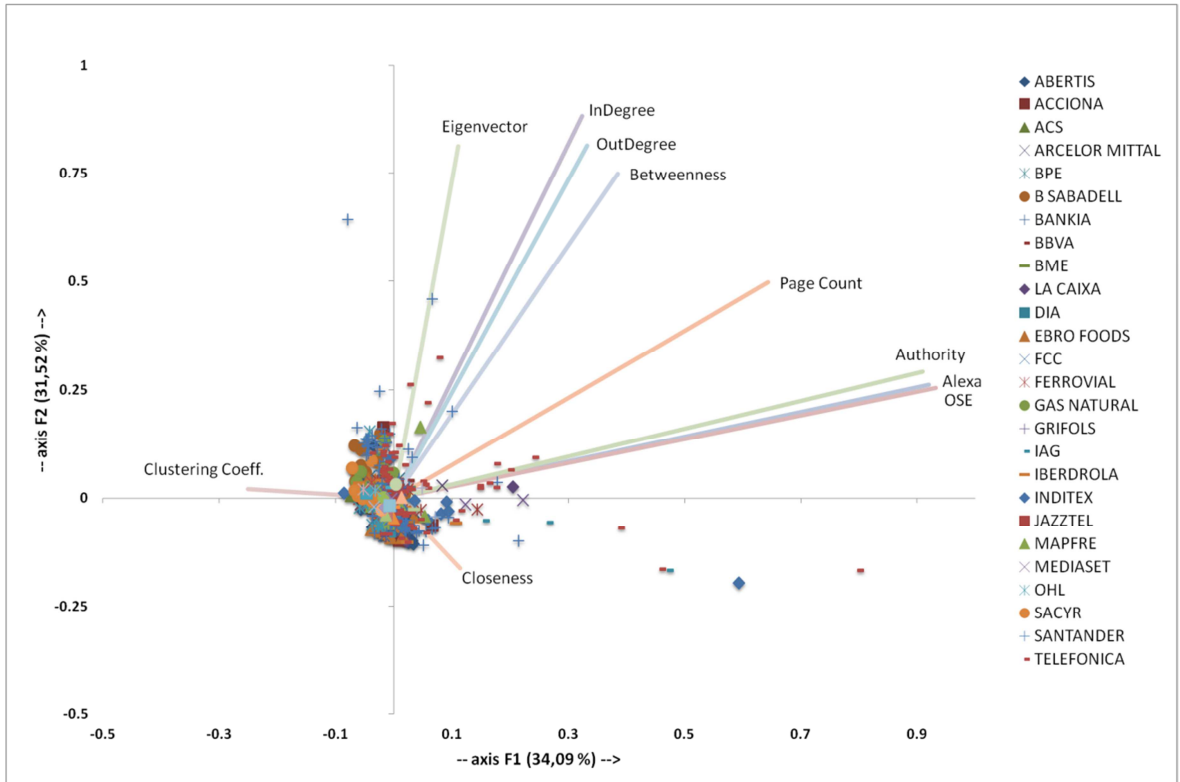

**Figure 3. Principal Component Analysis (PCA) of web impact and network indicators**